# ANTIPROLIFERATIVE MCR PEPTIDES BLOCK PHYSICAL INTERACTION OF INSULIN WITH RETINOBLASTOMA PROTEIN (RB) IN HUMAN LUNG CANCER CELLS

Razvan Tudor Radulescu[1]*  and  Kai Kehe[2]

[1]*Molecular Concepts Research (MCR), Munich, Germany;*

[2]*Institut für Pharmakologie & Toxikologie der Bundeswehr, 80937 Munich, Germany.*

*Corresponding author. E-mail: ratura@gmx.net

Keywords:   insulin, retinoblastoma protein (RB), hormone, growth factor, tumor suppressor, human lung cancer, MCR peptide, therapy, treatment, immunoprecipitation, magnetic beads, immunoblot

Running title: MCR peptides disrupt insulin-RB dimer in human lung cancer cells

**ABSTRACT**

Fifteen years ago, a structural analysis of the hormone insulin and the retinoblastoma tumor suppressor protein (RB) revealed that they may physically interact with one another. Subsequently, an RB peptide corresponding to the proposed RB binding site for insulin was found to recognize full-length insulin *in vitro*. As part of efforts aimed at developing this RB peptide into an anti-cancer drug, this molecule was chemically coupled to a cellular internalization signal and termed "MCR peptide". Meanwhile, several such MCR peptide variants have been demonstrated to restrain the proliferation of different human cancer cells *in vitro* and *in vivo*. Moreover, one of the MCR peptides coined MCR-10 was shown to be capable of interfering with the complex formation between insulin and RB in HepG2 human hepatoma cells, as monitored by immunofluorescence. This latter result indicating an *in vivo* association between insulin and RB was confirmed by a follow-up study combining the methods of co-immunoprecipitation and immunoblotting. Here, we provide evidence for the existence of the insulin-RB complex in A549 human non-small cell lung cancer cells. Specifically, we demonstrate this heterodimer by means of a magnetic beads-based immunoprecipitation approach and equally show that this dimer can be disrupted by MCR-4 or MCR-10 each of which is known to possess antiproliferative properties, yet to a much lesser extent by a control peptide. Thus, this investigation has yielded another important proof for the occurrence of the insulin-RB dimer and, furthermore, its validity as a target for antineoplastic MCR peptides.

**INTRODUCTION**

Over the past two decades, it has become clear that the healthy state of tissues is maintained by the functional balance between growth-promoting factors and tumor suppressor proteins. Among the latter, retinoblastoma protein (RB) plays a prominent role and, therefore, appears suitable to serve as a template for the design of anti-cancer drugs. In this context, MCR peptides have been developed to replace defective RB and thereby to even surpass its normal counterpart. Their active site is an RB fragment recognizing the hormone and growth factor insulin (1,2). MCR peptides have been shown to inhibit human cancer cell proliferation *in vitro* (3,4) and *in vivo* (5,6).

As previously shown in an immunofluorescence study for MCR-10, i.e. one of the members of the MCR peptide family (4), this type of compounds appears to achieve this effect, at least in part, by disrupting the physical interaction between insulin and RB, thus preventing that RB is inactivated by insulin. Since such interaction has also been demonstrated by sequential co-immunoprecipitation and immunoblotting (7), we have now chosen a magnetic beads-based variation of this approach to further investigate both the existence of the insulin-RB heterodimer and its possible disruption by two distinct MCR peptides as compared to a control peptide.

**MATERIALS AND METHODS**

*Peptides*. The amino acid sequences of the antiproliferative peptides MCR-4 and MCR-10, as synthesized and purified by Pichem (Graz, Austria) for this study, were previously reported (3). MCR-16 is the all-D isomer of the Antennapedia penetratin sequence (8) and was equally provided by Pichem.

*Tumor cells*. All experiments reported here were performed with the adherent human non-small cell lung cancer (NSCLC) cell line A549 known to be RB-positive.

*Basic experimental procedure*. After seeding of A549 cells cultured in RPMI/10% FCS into small-sized (T25) cell culture flasks with 2 million cells/flask and allowing the cells to adhere for 24 hours, the cells were incubated in the presence or absence of 50 μM MCR-4, MCR-10 or MCR-16 for 1 hour in one type of experiments and for 24 hours in another type of experiments. Each of these peptide treatments and the control treatment with RPMI was done in duplicate flasks. Subsequent to either peptide treatment, the cells were harvested by using a trypsin solution, lysed and processed as specified below.

*Magnetic beads-based co-immunoprecipitation of insulin and RB*. All the control- and peptide-treated cells were first washed with cold PBS and then harvested secondary to their trypsin-induced detachment from the flask walls. Subsequently, the collected cells were lysed in lysis buffer (10 mmol/l Tris, pH 8, 150 mmol/l NaCl, 1% NP-40, 0.5% sodium deoxycholate, 0.1% SDS, 1 mmol/l PMSF, 4 mg/ml aprotinin, 1 mmol/l sodium orthovanadate) for 15 min on ice and centrifuged at 4 degrees Celsius as well as maximum speed for another 15 min. The resulting supernatant was pre-cleared with Dynabeads protein A (Dynal Biotech ASA, Oslo, Norway) without antibody and then incubated with guinea pig polyclonal antibody A0564 specific for human insulin (DakoCytomation, Hamburg, Germany) that had previously been attached to Dynabeads protein A. The magnetically separated complex was washed several times with PBS and then boiled in SDS sample loading buffer for 5 min.

*Immunoblotting*. The protein samples captured by the magnetic beads-associated anti-insulin antibody were subjected to SDS-PAGE, blotted to nitrocellulose membranes and incubated with the mouse monoclonal antibody G3-245 directed against human RB (BD Pharmingen, Heidelberg, Germany). Signals were visualized by use of the Western Breeze Colorimetric kit (Invitrogen, Karlsruhe, FRG). The blot signals were scanned with the Fluor-S Imager and the respective bands were densitometrically quantitated with the software Quantity One (BioRad, Munich, Germany).

## RESULTS

We first investigated whether an incubation of A549 cells with MCR-4 for 1 hour is sufficiently long for this peptide to be able to interfere with insulin-RB complex formation as monitored by immunoblotting with the G3-245 antibody to RB subsequent to co-immunoprecipitation with the A0564 antibody to insulin (Fig. 1). The control lane on the left of Fig. 1 with no peptide added shows a strongly stained band on the height of the 110 kD RB band corresponding to the (lower) band on the right of Fig. 1 which represents an RB immunoblot without prior co-immunoprecipitation with the antibody to insulin, thus indicating that, in untreated and proliferating A549 cells, the insulin-RB physical interaction occurs to a significant degree. Furthermore, the RB band for the cells treated with MCR-4 is only faint (lane 2) whereas in those treated with the control peptide MCR-16 (lane 3) it has a rather moderate intensity, thus suggesting that MCR-4 interfered more with insulin binding to RB than MCR-16. In fact, a densitometric comparison of these bands revealed that this difference is significant (data not shown).

No MCRp    MCR-4    MCR-16

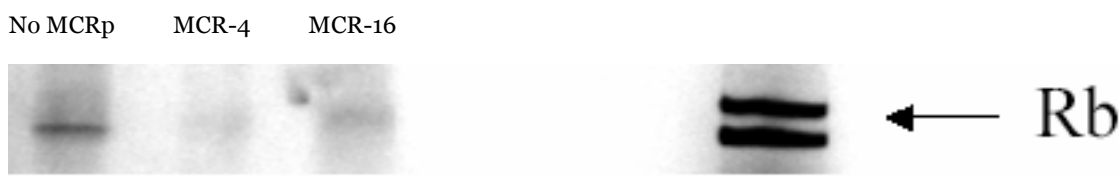

**Fig. 1** As can be seen from this RB immunoblot, MCR-4 interferes with insulin-RB complex formation in A549 cells more markedly than the control peptide MCR-16 after a 1-hour incubation.

When the duration of MCR-4 or MCR-16 treatment was increased from 1 hour to 24 hours, the results were similar, as revealed by the densitometric readout shown in Fig. 2.

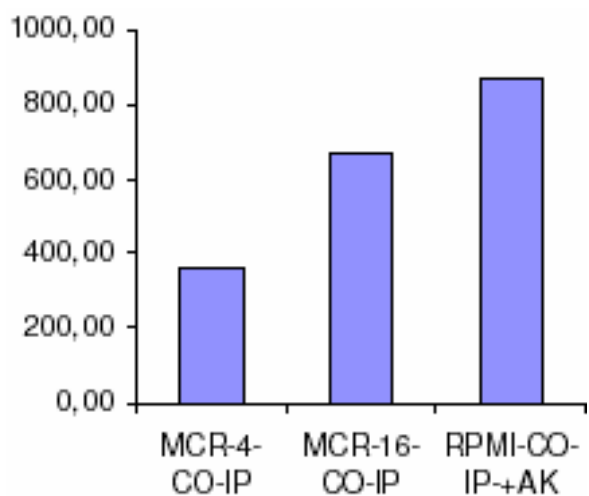

**Fig. 2** As can be gathered from this densitogram of an RB immunoblot, MCR-4 interferes with insulin-RB complex formation in A549 cells significantly more than the control peptide MCR-16 after a 24-hour incubation.

The disruption of the insulin-RB complex was equally observed when MCR-10 was used (Fig. 3, lane 2). In this experiment, MCR-16 has hardly influenced the occurrence of insulin-RB dimer formation and thus could not prevent the apparition of a clearcut RB band (Fig. 3, lane 3).

No MCRp    MCR-10    MCR-16  No anti-Ins  Ctrl RB WB

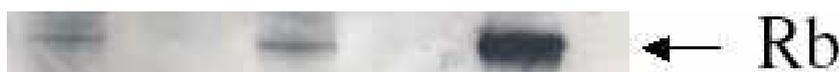

**Fig. 3** This RB immunoblot reveals that MCR-10 interferes with the physical interaction between insulin and RB in A549 cells considerably more than the control peptide MCR-16 after a 24-hour incubation.

As an additional control to those cells not treated with any peptide, yet subjected to the procedure of co-immunoprecipitation with the antibody to insulin and subsequent immunoblotting with the antibody to RB (Fig. 3, lane 1) as well as other cells directly subjected to RB immunoblotting without prior insulin co-immunoprecipitation (Fig. 3, lane 5), we also included a condition (Fig. 3, lane 4) in which the antibody to insulin had been left out from the combined immunoprecipitation/RB immunoblot procedure to check for the presence of any non-specific matrix/magnetic bead effects. This type of RB immunoblot yielded no visible band (Fig. 3, lane 4) which upon densitometric analysis corresponded to an only minimal signal (data not shown).

**DISCUSSION**

The purpose of the present study has been to investigate if the range of cells in which one can observe insulin-RB complexes can be extended from human HepG2 hepatoma cells (4,7) to A549 human non-small cell lung cancer (NSCLC) cells and, provided that such interactions can be recorded in these cells, whether they can be specifically disturbed by means of MCR peptides, as initially shown for MCR-10 in HepG2 cells (4).

Our present results clearly indicate that the insulin-RB heterodimer exists to a significant degree also in proliferating A549 cells and, moreover, that any of the antiproliferative MCR peptides investigated here, i.e. MCR-4 and MCR-10, can interfere with the formation of this protein complex in this experimental system. Notably, each of these peptides has previously been shown to potently block the proliferation of A549 cells *in vitro* (3) and, furthermore, MCR-4 was shown to exert *in vivo* anti-cancer activity against A549 xenografts in nude mice (5). As such, several major conclusions should be drawn.

Firstly, the RB immunoblot band obtained after initial co-immunoprecipitation with the antibody to insulin and running at about 110 kD was much slimmer than the wide band located in an approximate molecular weight range of 110-115 kD and showing up in the control RB immunoblot performed without prior insulin co-immunoprecipitation. This suggests that, similar to some viral oncoproteins and a number of cellular ligands, insulin binds only to a subset of RB isoforms, most likely to the hypophosphorylated and thus active RB form. Future studies will therefore have to investigate the validity of this assumption, e.g. by means of an antibody specific for this RB isoform.

Secondly, it seems that the insulin-RB complex is more stable and perhaps also more decisive for positive cell growth regulation than the previously reported dimer between human papilloma virus 16 E7 protein and RB since the latter dimer was observed *in vitro*, yet, interestingly, not *in vivo*, specifically not in several cervical carcinoma cell lines (9). Beyond the possible explanation of an increased and accelerated (ubiquitin-linked) degradation of E7-RB complexes, it is tempting to speculate that the growth stimulation and neoplastic transformation commonly attributed to E7 may actually, at least in part, be mediated by growth factors such as insulin. As such, insulin may be upregulated in response to E7, similar to the IGF-2 induction as part of the SV40 large T viral oncoprotein-driven carcinogenesis in experimental pancreatic cancer (10), and subsequently such overexpressed insulin could transduce E7-initiated viral oncogenesis by binding and thereby inactivating RB. Such insulin positivity of cancer cells may be highly predictive of a state of malignancy, for instance of cervix carcinoma where it may parallel IGF-2 expression which has been found to be clearly superior to HPV positivity as a biomarker for early diagnosis (11) and/or metastatic spread to lymph nodes, in particular within so-called pauci-cellular metastases (12). Consequently, targeting insulin by means of antiproliferative MCR peptides in order to liberate RB from the insulin-RB complex should become a successful approach in the treatment of cancer including those malignancies associated with oncoviral proteins such as E7.

Thirdly, towards attempting to minimize the development of any potential cancer cell resistance to therapy, it should be specified that, based on the results of this study, MCR-4 appears to have at least two distinct mechanisms of action and should therefore be a particularly suitable drug candidate. As such, MCR-4 not only interferes with the physical interaction between insulin and RB, as demonstrated here, but has also been shown to induce a high intracellular level of the cyclin-dependent kinase (cdk) inhibitor p21 in A549 cells that was found to coincide with their rapid apoptosis (13). Since this cell death occurred in the absence of any apparent caspase activation, this MCR-4 mechanism is thus devoid of yet another potential cause for drug resistance.

Taken together, our present data support the notion of a potential involvement of the insulin-RB complex in cancer pathophysiology and, moreover, warrant further accelerated development of MCR peptides, in particular of MCR-4, into clinically applicable anti-cancer drugs.

**ACKNOWLEDGMENTS**

We thank Mina Zahrabi for excellent technical assistance.